\documentclass[]{aastex631}
\pdfoutput=1 

\graphicspath{{./}{figures/}}
\usepackage[T5,T1]{fontenc}
\begin{document}

\title{Search for Astrophysical Neutrinos from 1FLE Blazars with IceCube}
\affiliation{III. Physikalisches Institut, RWTH Aachen University, D-52056 Aachen, Germany}
\affiliation{Department of Physics, University of Adelaide, Adelaide, 5005, Australia}
\affiliation{Dept. of Physics and Astronomy, University of Alaska Anchorage, 3211 Providence Dr., Anchorage, AK 99508, USA}
\affiliation{Dept. of Physics, University of Texas at Arlington, 502 Yates St., Science Hall Rm 108, Box 19059, Arlington, TX 76019, USA}
\affiliation{CTSPS, Clark-Atlanta University, Atlanta, GA 30314, USA}
\affiliation{School of Physics and Center for Relativistic Astrophysics, Georgia Institute of Technology, Atlanta, GA 30332, USA}
\affiliation{Dept. of Physics, Southern University, Baton Rouge, LA 70813, USA}
\affiliation{Dept. of Physics, University of California, Berkeley, CA 94720, USA}
\affiliation{Lawrence Berkeley National Laboratory, Berkeley, CA 94720, USA}
\affiliation{Institut f{\"u}r Physik, Humboldt-Universit{\"a}t zu Berlin, D-12489 Berlin, Germany}
\affiliation{Fakult{\"a}t f{\"u}r Physik {\&} Astronomie, Ruhr-Universit{\"a}t Bochum, D-44780 Bochum, Germany}
\affiliation{Universit{\'e} Libre de Bruxelles, Science Faculty CP230, B-1050 Brussels, Belgium}
\affiliation{Vrije Universiteit Brussel (VUB), Dienst ELEM, B-1050 Brussels, Belgium}
\affiliation{Department of Physics and Laboratory for Particle Physics and Cosmology, Harvard University, Cambridge, MA 02138, USA}
\affiliation{Dept. of Physics, Massachusetts Institute of Technology, Cambridge, MA 02139, USA}
\affiliation{Dept. of Physics and The International Center for Hadron Astrophysics, Chiba University, Chiba 263-8522, Japan}
\affiliation{Department of Physics, Loyola University Chicago, Chicago, IL 60660, USA}
\affiliation{Dept. of Physics and Astronomy, University of Canterbury, Private Bag 4800, Christchurch, New Zealand}
\affiliation{Dept. of Physics, University of Maryland, College Park, MD 20742, USA}
\affiliation{Dept. of Astronomy, Ohio State University, Columbus, OH 43210, USA}
\affiliation{Dept. of Physics and Center for Cosmology and Astro-Particle Physics, Ohio State University, Columbus, OH 43210, USA}
\affiliation{Niels Bohr Institute, University of Copenhagen, DK-2100 Copenhagen, Denmark}
\affiliation{Dept. of Physics, TU Dortmund University, D-44221 Dortmund, Germany}
\affiliation{Dept. of Physics and Astronomy, Michigan State University, East Lansing, MI 48824, USA}
\affiliation{Dept. of Physics, University of Alberta, Edmonton, Alberta, Canada T6G 2E1}
\affiliation{Erlangen Centre for Astroparticle Physics, Friedrich-Alexander-Universit{\"a}t Erlangen-N{\"u}rnberg, D-91058 Erlangen, Germany}
\affiliation{Physik-department, Technische Universit{\"a}t M{\"u}nchen, D-85748 Garching, Germany}
\affiliation{D{\'e}partement de physique nucl{\'e}aire et corpusculaire, Universit{\'e} de Gen{\`e}ve, CH-1211 Gen{\`e}ve, Switzerland}
\affiliation{Dept. of Physics and Astronomy, University of Gent, B-9000 Gent, Belgium}
\affiliation{Dept. of Physics and Astronomy, University of California, Irvine, CA 92697, USA}
\affiliation{Karlsruhe Institute of Technology, Institute for Astroparticle Physics, D-76021 Karlsruhe, Germany }
\affiliation{Karlsruhe Institute of Technology, Institute of Experimental Particle Physics, D-76021 Karlsruhe, Germany }
\affiliation{Dept. of Physics, Engineering Physics, and Astronomy, Queen's University, Kingston, ON K7L 3N6, Canada}
\affiliation{Dept. of Physics and Astronomy, University of Kansas, Lawrence, KS 66045, USA}
\affiliation{Department of Physics and Astronomy, UCLA, Los Angeles, CA 90095, USA}
\affiliation{Centre for Cosmology, Particle Physics and Phenomenology - CP3, Universit{\'e} catholique de Louvain, Louvain-la-Neuve, Belgium}
\affiliation{Department of Physics, Mercer University, Macon, GA 31207-0001, USA}
\affiliation{Dept. of Astronomy, University of Wisconsin{\textendash}Madison, Madison, WI 53706, USA}
\affiliation{Dept. of Physics and Wisconsin IceCube Particle Astrophysics Center, University of Wisconsin{\textendash}Madison, Madison, WI 53706, USA}
\affiliation{Institute of Physics, University of Mainz, Staudinger Weg 7, D-55099 Mainz, Germany}
\affiliation{Department of Physics, Marquette University, Milwaukee, WI, 53201, USA}
\affiliation{Institut f{\"u}r Kernphysik, Westf{\"a}lische Wilhelms-Universit{\"a}t M{\"u}nster, D-48149 M{\"u}nster, Germany}
\affiliation{Bartol Research Institute and Dept. of Physics and Astronomy, University of Delaware, Newark, DE 19716, USA}
\affiliation{Dept. of Physics, Yale University, New Haven, CT 06520, USA}
\affiliation{Dept. of Physics, University of Oxford, Parks Road, Oxford OX1 3PU, UK}
\affiliation{Dept. of Physics, Drexel University, 3141 Chestnut Street, Philadelphia, PA 19104, USA}
\affiliation{Physics Department, South Dakota School of Mines and Technology, Rapid City, SD 57701, USA}
\affiliation{Dept. of Physics, University of Wisconsin, River Falls, WI 54022, USA}
\affiliation{Dept. of Physics and Astronomy, University of Rochester, Rochester, NY 14627, USA}
\affiliation{Department of Physics and Astronomy, University of Utah, Salt Lake City, UT 84112, USA}
\affiliation{Oskar Klein Centre and Dept. of Physics, Stockholm University, SE-10691 Stockholm, Sweden}
\affiliation{Dept. of Physics and Astronomy, Stony Brook University, Stony Brook, NY 11794-3800, USA}
\affiliation{Dept. of Physics, Sungkyunkwan University, Suwon 16419, Korea}
\affiliation{Institute of Basic Science, Sungkyunkwan University, Suwon 16419, Korea}
\affiliation{Institute of Physics, Academia Sinica, Taipei, 11529, Taiwan}
\affiliation{Dept. of Physics and Astronomy, University of Alabama, Tuscaloosa, AL 35487, USA}
\affiliation{Dept. of Astronomy and Astrophysics, Pennsylvania State University, University Park, PA 16802, USA}
\affiliation{Dept. of Physics, Pennsylvania State University, University Park, PA 16802, USA}
\affiliation{Dept. of Physics and Astronomy, Uppsala University, Box 516, S-75120 Uppsala, Sweden}
\affiliation{Dept. of Physics, University of Wuppertal, D-42119 Wuppertal, Germany}
\affiliation{DESY, D-15738 Zeuthen, Germany}

\author[0000-0001-6141-4205]{R. Abbasi}
\affiliation{Department of Physics, Loyola University Chicago, Chicago, IL 60660, USA}

\author[0000-0001-8952-588X]{M. Ackermann}
\affiliation{DESY, D-15738 Zeuthen, Germany}

\author{J. Adams}
\affiliation{Dept. of Physics and Astronomy, University of Canterbury, Private Bag 4800, Christchurch, New Zealand}

\author[0000-0003-2252-9514]{J. A. Aguilar}
\affiliation{Universit{\'e} Libre de Bruxelles, Science Faculty CP230, B-1050 Brussels, Belgium}

\author[0000-0003-0709-5631]{M. Ahlers}
\affiliation{Niels Bohr Institute, University of Copenhagen, DK-2100 Copenhagen, Denmark}

\author{M. Ahrens}
\affiliation{Oskar Klein Centre and Dept. of Physics, Stockholm University, SE-10691 Stockholm, Sweden}

\author[0000-0002-9534-9189]{J.M. Alameddine}
\affiliation{Dept. of Physics, TU Dortmund University, D-44221 Dortmund, Germany}

\author{A. A. Alves Jr.}
\affiliation{Karlsruhe Institute of Technology, Institute for Astroparticle Physics, D-76021 Karlsruhe, Germany }

\author{N. M. Amin}
\affiliation{Bartol Research Institute and Dept. of Physics and Astronomy, University of Delaware, Newark, DE 19716, USA}

\author{K. Andeen}
\affiliation{Department of Physics, Marquette University, Milwaukee, WI, 53201, USA}

% ...

\collaboration{383}{IceCube Collaboration}

\begin{abstract}
\replaced{The origin of astrophysical neutrinos has yet to be determined.}{The majority of astrophysical neutrinos have undetermined origins.} The IceCube Neutrino Observatory has observed astrophysical neutrinos but has not yet identified their sources. Blazars are promising source candidates, but previous searches for neutrino emission from populations of blazars detected in $\gtrsim$ GeV gamma-rays have not observed any significant neutrino excess. Recent findings in multi-messenger astronomy indicate that high-energy photons, co-produced with high-energy neutrinos, are likely to be absorbed and reemitted at lower energies. Thus, lower-energy photons may be better indicators of TeV-PeV neutrino production. This paper presents the first time-integrated stacking search for astrophysical neutrino emission from MeV-detected blazars in the first \textit{Fermi}-LAT low energy catalog (1FLE) using ten years of IceCube muon-neutrino data. The results of this analysis are found to be consistent with a background-only hypothesis. Assuming an E$^{-2}$ neutrino spectrum and proportionality between the blazars’ MeV gamma-ray fluxes and TeV-PeV neutrino flux, the upper limit on the 1FLE blazar energy-scaled neutrino flux is determined to be $1.64 \times 10^{-12}$ TeV cm$^{-2}$ s$^{-1}$ at 90\% confidence level. This upper limit is approximately $1\%$ of IceCube's diffuse muon-neutrino flux measurement.
\end{abstract}

\section{\label{sec:Intro}Introduction}
High-energy cosmic rays have been observed arriving at Earth \citep{Zyla:2020zbs,PierreAuger:2021hun,ABBASI2016131}, but their origins have yet to be determined. The observation of these particles with PeV and higher energies implies the existence of powerful astrophysical particle accelerators; however, the nature and locations of these objects remains uncertain. Charged particles with energies smaller than $\sim10^{19}$ eV traveling through the Universe are deflected by interstellar and intergalactic magnetic fields \citep{PhysRevD.104.083017}, thereby making localization of cosmic ray sources difficult. 

In contrast to charged particles, photons and neutrinos travel along a straight path from their sources making them useful messengers for observing cosmic ray accelerators. Unlike photons which can be produced by leptonic or hadronic processes, production of a large neutrino flux requires the presence of a hadronic component. Therefore, the emission of neutrinos is generally considered a strong sign of cosmic ray acceleration. Additionally, photon attenuation probability increases rapidly with energy $\gtrsim$ 30 GeV over cosmic distances \citep{2017ApJS..232...18A}. On the other hand, the neutrino's small interaction cross-section allows it to travel great distances without absorption making them effective at probing distances that photons cannot.

These same properties necessitate a large detector volume to effectively observe neutrinos, especially those of astrophysical origin which have a relatively small flux compared to solar and atmospheric neutrinos. The IceCube Neutrino Observatory (hereafter, IceCube) is a cubic-kilometer Cherenkov light detector located at the geographic South Pole \citep{2017JInst..12P3012A}. In 2013, IceCube discovered the existence of a flux of high-energy, astrophysical neutrinos \citep{2013Sci...342E...1I}. No sources of these astrophysical neutrinos have yet been \replaced{definitively discovered.}{found with post-trial significance exceeding 5$\sigma$.}

Active galactic nuclei (AGN) are highly energetic objects which have long been theorized to be potential sites for particle acceleration and neutrino production \citep{1997ApJ...488..669H}. Accretion of surrounding matter onto a super-massive black hole creates an ideal environment for acceleration and interaction of cosmic particles. In some AGN, particles may additionally be accelerated along a narrow, relativistic jet. Four AGN make up the most significant combined group of sources in an analysis of 10 years of IceCube data \citep{2020PhRvL.124e1103A}; among these are the Seyfert and starburst galaxy NGC 1068 and blazar TXS 0506+056.

This blazar, TXS 0506+056, is the first object to show evidence for astrophysical neutrino emission \citep{2018Sci...361.1378I,2018Sci...361..147I} \added{with greater than 3$\sigma$ significance}. Blazars are AGN which have a relativistic jet directed toward Earth, providing a potentially greater observed flux of accelerated particles \citep{1995PASP..107..803U}. Current observations of blazars are unable to distinguish between leptonic, hadronic, and mixed emission scenarios \citep{2019MNRAS.483L..12C,B_ttcher_2013}. When a hadronic component is included, a neutrino counterpart is expected alongside photon production; hence, detections of neutrinos from blazars can help characterize their makeup. 

Previous IceCube analyses \citep{2017ApJ...835...45A,2019ICRC...36..916H}, though, have not found significant astrophysical neutrino emission from blazars in the 2 year \textit{Fermi}-LAT AGN catalog (2LAC) \citep{2011ApJ...743..171A} and the \textit{Fermi}-LAT third high energy source catalog (3FHL) \citep{2017ApJS..232...18A}. Recent theoretical hypotheses \citep{2016PhRvL.116g1101M} suggest that photons produced alongside astrophysical neutrinos in cosmic accelerators will be more effectively observed at reduced energies due to processes which attenuate high-energy photons. At the source, high-energy gamma-rays are unlikely to escape their environment without being absorbed, for example by pair-production with other gamma-rays, inverse Compton scattering, or synchrotron cooling \citep{1991A&A...251..723M,PhysRevD.63.023003}. Photons that do escape their source with $\gtrsim$ GeV energies have a high probability of cascading to lower energies via interaction with extragalactic background light and the cosmic microwave background as they travel from their source \citep{1975Ap&SS..32..461B}. These processes lead not only to an increased observed electromagnetic flux at MeV energies but also to the possibility for some neutrino sources to be unobserved in the GeV-TeV energy range \citep{2016PhRvL.116g1101M}. In addition, blazar spectral energy distributions tend to peak in the MeV regime, especially for the most luminous objects \citep{1998MNRAS.299..433F,Gao_2017,Ojha2019Neutrinos}. For these reasons, MeV photons could be more effective indicators of TeV-PeV neutrinos.

\section{\label{sec:sources}Sources}
\begin{figure}[!htb]
    \epsscale{0.75}
    \plotone{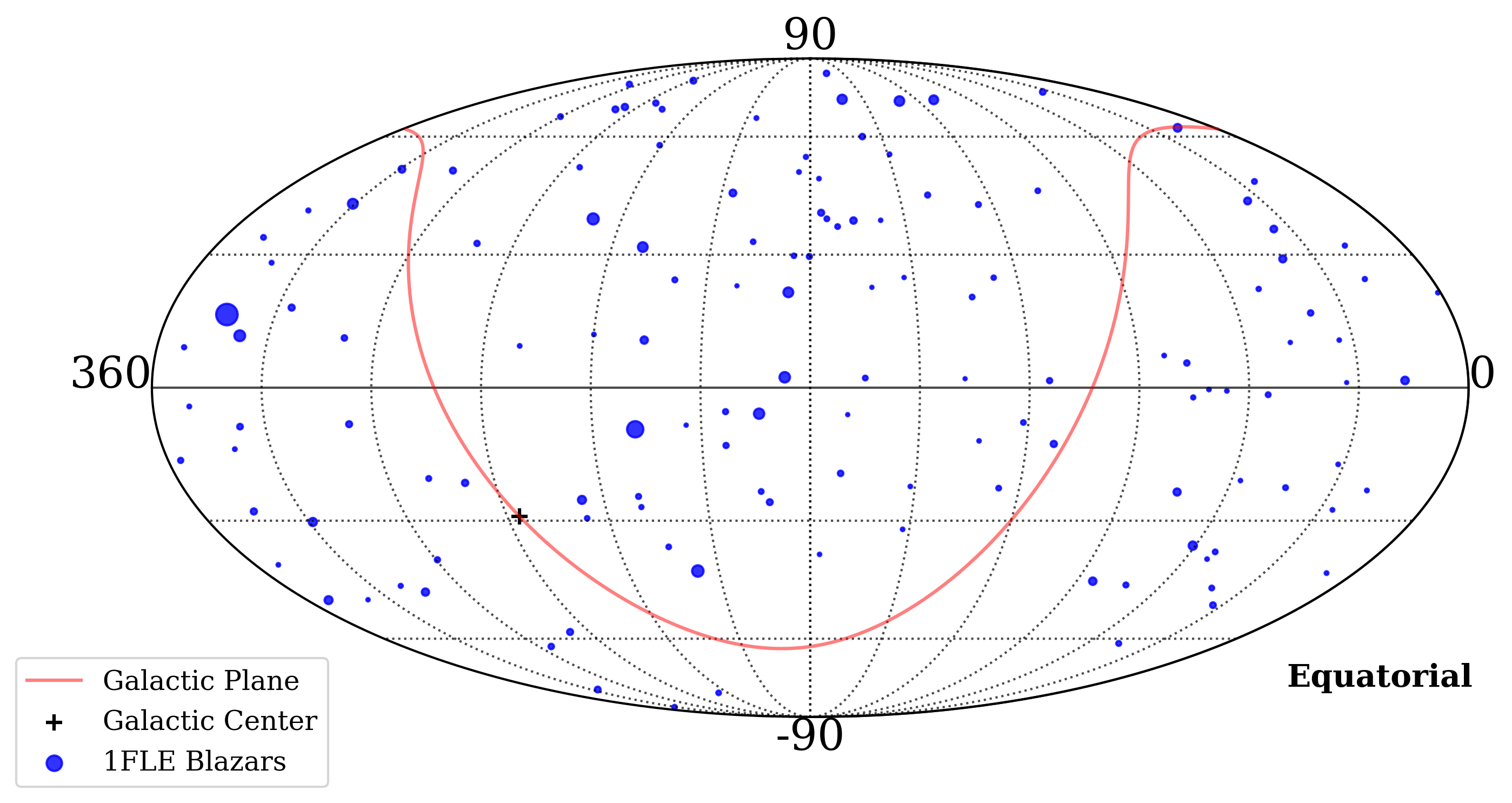}
    \caption{Sky map showing locations of 1FLE blazars used as sources in this analysis in equatorial coordinates. The marker size for each blazar is proportional to its MeV gamma-ray flux. Also shown are the Galactic center (black cross) and Galactic plane (red line) for reference. Note that despite some sources' proximity to the Galactic plane, all blazars are extragalactic. \label{fig:map}}
\end{figure}

This analysis searches for correlation between IceCube neutrinos and 137 blazars in the first catalog of \textit{Fermi}-LAT sources detected below 100 MeV (1FLE) \citep{2018A&A...618A..22P}. These blazars are mapped in equatorial coordinates in Figure \ref{fig:map} with marker sizes proportional to each blazar's respective 30-100 MeV photon energy flux. The Large Area Telescope (LAT) on board the Fermi Gamma-ray Space Telescope (\textit{Fermi}) observes the Universe in gamma-rays of energies from 20 MeV to over 300 GeV, its most sensitive energies falling in the GeV band \citep{2009ApJ...697.1071A}. The 1FLE catalog was created using a wavelet transform method to localize sources and fit for fluxes which is described in detail in \citet{2018A&A...618A..22P}. The blazars used in this analysis are classified as either flat-spectrum radio quasars (FSRQs) or BL Lacertae objects (BL Lacs). These classifications are obtained from each blazar's respective source association in the 10-year \textit{Fermi}-LAT data release (4FGL-DR2) \citep{Abdollahi_2020} in order to have the most updated identification. It is important to note, however, that blazars are not split into separate classifications for use in this analysis. While there are blazars shared between the 1FLE catalog and the catalogs used in previous IceCube blazar stacking analyses (2LAC and 3FHL), the number of sources and the photon flux energy range differentiate this analysis from previous, similar ones. A brief overview of these quantities in this analysis and those previously mentioned is presented in Table \ref{tab:comp}. 

The 30-100 MeV photon energy range of the 1FLE catalog is below the peak sensitivity of the \textit{Fermi}-LAT instrument. In general, this leads to a lower precision on measured quantities. For instance, the 1FLE catalog localizes sources within error radii of about $0.25^{\circ}$ \citep{2018A&A...618A..22P}. In comparison, 4FGL sources, which are localized using photons where \textit{Fermi}-LAT is more sensitive above 100 MeV, have error radii that tend to be around $0.05^{\circ}$ (varying up to about $0.5^{\circ}$ and down to about $0.005^{\circ}$) \citep{Abdollahi_2020}. Despite these greater uncertainties in the 1FLE catalog, the locations of the blazars do not vary greatly from their 4FGL associations. Additionally, the 1FLE source detection methods are designed to detect faint sources while also limiting source confusion (see \citet{2018A&A...618A..22P} for full explanation). This specific focus of the 1FLE catalog on detection efficiency in the 30-100 MeV photon energy range makes it a good choice for this analysis. Of further importance is the fact that the source localization errors are still smaller or similar to the IceCube pointing resolution, so they do not contribute significant error when correlating with IceCube events. All this considered, without instruments that are sensitive to the MeV range of the electromagnetic spectrum, the 1FLE catalog acts as an important bridge over previously unexplored energies. A full list of source blazars used in this analysis and relevant quantities can be found in Appendix \ref{app:sl}. 

\section{\label{sec:data}Data}
The IceCube detector \citep{2017JInst..12P3012A} consists of 5160 digital optical modules (DOMs) arranged in a cubic kilometer array of Antarctic ice. The main component of each DOM is a photo-multiplier tube which collects the light created by relativistic particles traveling through the ice. These particles are produced as byproducts of neutrino interactions and emit Cherenkov radiation when traveling faster than the local speed of light through the ice. IceCube can detect all flavors of neutrinos, but no distinction can be made between neutrino ($\nu$) and anti-neutrino ($\bar{\nu}$) except in rare cases such as the Glashow resonance \citep{2021Natur.591..220I}. IceCube observes two main event topologies at energies $>100$ GeV: cascades and tracks. Cascade events appear as a near-spherical propagation of light resulting from neutral-current (NC) neutrino interactions of all flavors or from charged-current (CC) interactions of electron- and tau-neutrinos ($\nu_e\bar{\nu}_e$ and $\nu_{\tau}\bar{\nu}_{\tau}$, respectively). At energies greater than a few hundred TeV, the CC $\nu_{\tau}\bar{\nu}_{\tau}$ interaction can produce a double-cascade light signature as the resulting tau particle travels a distance resolvable by IceCube \citep{PhysRevD.93.022001}. The data used in this analysis are track events. The track topology mostly results from the CC interactions of muon-neutrinos ($\nu_{\mu}+\bar{\nu}_{\mu}$); the muon produced in these interactions travels of order several kilometers through the Antarctic ice as it loses energy, thereby depositing light in a linear pattern. This allows for tracks to be reconstructed with angular resolution of $\lesssim 1^{\circ}$ for neutrinos with energies  $\gtrsim$ 1 TeV \citep{2020PhRvL.124e1103A}. This angular resolution is smaller than that of cascade events which makes track events favored for point source searches. 

The dataset used to perform this analysis contains 1134451 IceCube track events collected between April 6, 2008 and July 8, 2018 from the entire sky. As the estimated angular uncertainty for each individual event does not include systematic uncertainties, a minimum angular uncertainty of $0.2^{\circ}$ is applied \citep{2020PhRvL.124e1103A}. A majority of this dataset is background events created by atmospheric neutrinos and muons resulting from cosmic ray interactions with Earth's atmosphere. Significant reduction of this background is achieved in the Northern hemisphere due to shielding from muons by the Earth. Further details about the dataset and event selection methods may be found in \citet{2021arXiv210109836I}.

\section{\label{sec:analysis}Analysis methods}
This analysis uses a well-defined \citep{2017ApJ...835...45A,2020ApJ...898..117A,BRAUN2008299} unbinned, maximum likelihood method which makes use of location and energy information to determine possible correlation between IceCube neutrinos and 1FLE blazars. It is a stacked search which looks for cumulative neutrino emission from the source list as a whole, rather than individually. This time-independent analysis assumes steady emission from the sources as opposed to searching for transient emission. The likelihood, $\mathcal{L}$---which is a function of two free variables, the number of signal events, $n_s$, and the neutrino spectral index, $\gamma$---is defined in Equation \ref{eqn:lh}. This likelihood can then be maximized to obtain a best-fit $n_s$ and $\gamma$.
\begin{equation}\label{eqn:lh}
    \mathcal{L}(n_s , \gamma)= \prod_i^N \left( \frac{n_s}{N} S_i + \left( 1 - \frac{n_s}{N} \right) B_i \right)
\end{equation}
The value $N$ is the total number of events in the data sample. The likelihood consists of two PDFs for each event, $i$: one for signal, $S_i$, and one for background, $B_i$. The PDFs each consist of a spatial and an energy factor which correlate each event with a source. 

As this is a stacked search, $S_i$ is the weighted sum over every source $j$ such that $S_i = \sum_j^M w_j R_j S_i^j$ for $M$ total sources. The weight of each source consists of a detector acceptance weight, $R_j$, based on the effective area at the source location and a source weighting scheme, $w_j$, based on a chosen emission hypothesis. These weights are normalized such that $\sum_j^M w_j R_j = 1$; for more details see \citet{2020ApJ...898..117A}. The two source weighting schemes chosen for use in this analysis are equally-weighted sources (referred to as equal weights), and sources weighted by their integrated photon energy fluxes in the 30-100 MeV range (referred to as flux weights).
\begin{deluxetable*}{cccccc}
    \tablenum{1}
    \tablecolumns{10}
    \tablecaption{Comparison of IceCube blazar stacking analyses. \label{tab:comp}}
    \tablewidth{\textwidth}
    \tablehead{
    \colhead{Analysis} & \colhead{Source Catalog} & \colhead{Total Number of Blazars} & \colhead{Flux Weight Energy Range} \\
    \colhead{} & \colhead{} & \colhead{}  & \colhead{(GeV)}
    }
        \startdata
        \citet{2017ApJ...835...45A} & 2LAC & 862 & 0.1 - 100 \\
        \citet{2019ICRC...36..916H} & 3FHL & \replaced{1301}{745} & --- \\
        This work & 1FLE & 137 & 0.03 - 0.1
        \enddata
    \tablecomments{Overview of two blazar stacking analyses is shown in addition to this work. In comparison to the 137 1FLE blazars studied in this analysis, 862 blazars from the 2LAC \citep{2011ApJ...743..171A} catalog were analyzed with flux weights in the 0.1 - 100 GeV energy range, and \replaced{1301}{745} blazars from the 3FHL \citep{2017ApJS..232...18A} catalog were analyzed without flux weights.}
\end{deluxetable*}

A test statistic (TS) is then constructed, as in Equation \ref{eqn:ts}, to compare the best-fit signal hypothesis to a background-only hypothesis that is characterized by having zero signal events. This TS is used to determine the significance of the correlation in the form of a p-value, defined as the probability that the observed TS is a result of background fluctuations.
\begin{equation}\label{eqn:ts}
     \text{TS} = -2 \log \left( \frac{\mathcal{L}(n_s=0)}{\mathcal{L}(n_s , \gamma)}\right)
\end{equation}

As mentioned in Section \ref{sec:data}, the data sample is background-dominated. IceCube's location at the geographic South Pole leads to a declination-dependent detector acceptance and event rate. Therefore, a model-independent background estimation is obtained by randomizing the right ascension of data events. This background estimation is repeated many times to create a distribution to which the observed TS may be compared in order to determine the p-value. 

\section{\label{sec:res}Results}
\begin{figure}[!htb]
    \epsscale{0.75}
    \plotone{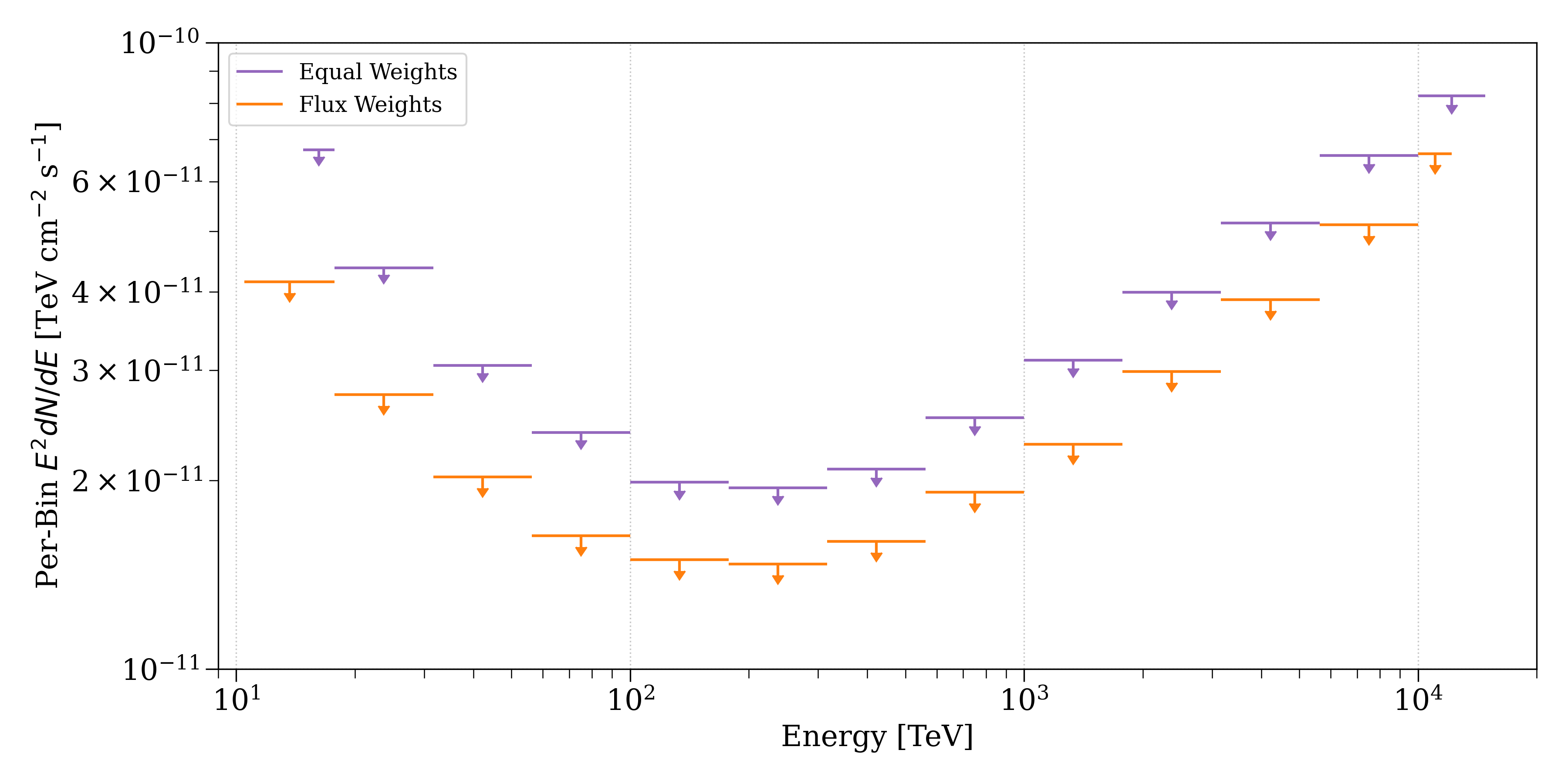}
    \caption{90\% confidence level upper limits on energy-scaled neutrino ($\nu_{\mu}+\bar{\nu}_{\mu}$) flux from 1FLE blazars in four differential bins per decade of energy. These limits correspond to an E$^{-2}$ neutrino spectrum within each energy bin and do not include systematic uncertainties. The overall energy range is that which contributed 90\% of the total sensitivity for each respective weighting scheme. \label{fig:DS}}
\end{figure}

The best-fit number of signal events for both source weighting hypotheses is found to be zero, hence our observations are consistent with the background-only hypothesis. Upper limits (ULs) are set on the energy-scaled neutrino flux ($E^2 dN/dE$) for various, assumed neutrino spectral indices and do not include systematic uncertainties. Such systematic uncertainties arise from DOM efficiency, absorption or scattering of photons in the ice, and photo-nuclear interaction models; the total systematic uncertainty on neutrino fluxes is approximately 11\% \citep{Aartsen_2017_7yr}. The results of this analysis are summarized in Table \ref{tab:res}. Additionally, Figure \ref{fig:DS} shows differential ULs on energy-scaled neutrino flux for an $E^{-2}$ neutrino spectrum in bins of energy. The range of energies shown in the figure is that which contributes 90\% of the integrated sensitivity for each weighting scheme.
\begin{deluxetable*}{ccccccc}
    \tablenum{2}
    \tablecolumns{10}
    \tablecaption{Stacking analysis results for 1FLE blazars. \label{tab:res}}
    \tablewidth{\textwidth}
    \tablehead{
    \colhead{Weighting} & \colhead{TS} & \colhead{n$_s$} & \multicolumn{4}{c}{Energy-scaled Neutrino Flux UL (TeV cm$^{-2}$ s$^{-1}$ at $1$ TeV)} \\
    \colhead{Scheme} & \colhead{} & \colhead{} & \colhead{$\gamma=2.0$} & \colhead{$\gamma=2.37$} & \colhead{$\gamma=2.5$} & \colhead{$\gamma=3.0$}
    }
        \startdata
        Equal & 0 & 0 & $2.27 \times 10^{-12}$ & $1.48 \times 10^{-11}$ & $2.48 \times 10^{-11}$ & $9.72 \times 10^{-11}$ \\
        Flux & 0 & 0 & $1.64 \times 10^{-12}$ & $9.98 \times 10^{-12}$ & $1.65 \times 10^{-11}$ & $6.07 \times 10^{-11}$
        \enddata
    \tablecomments{The 90\% confidence level upper limits on stacked, energy-scaled neutrino ($\nu_{\mu}+\bar{\nu}_{\mu}$) flux are shown for several assumed spectral indices at a reference energy of 1 TeV, along with the best-fit TS and n$_s$, for each weighting scheme. These upper limits do not include systematic uncertainties.}
\end{deluxetable*}

\begin{figure}[!htb]
    \epsscale{0.75}
    \plotone{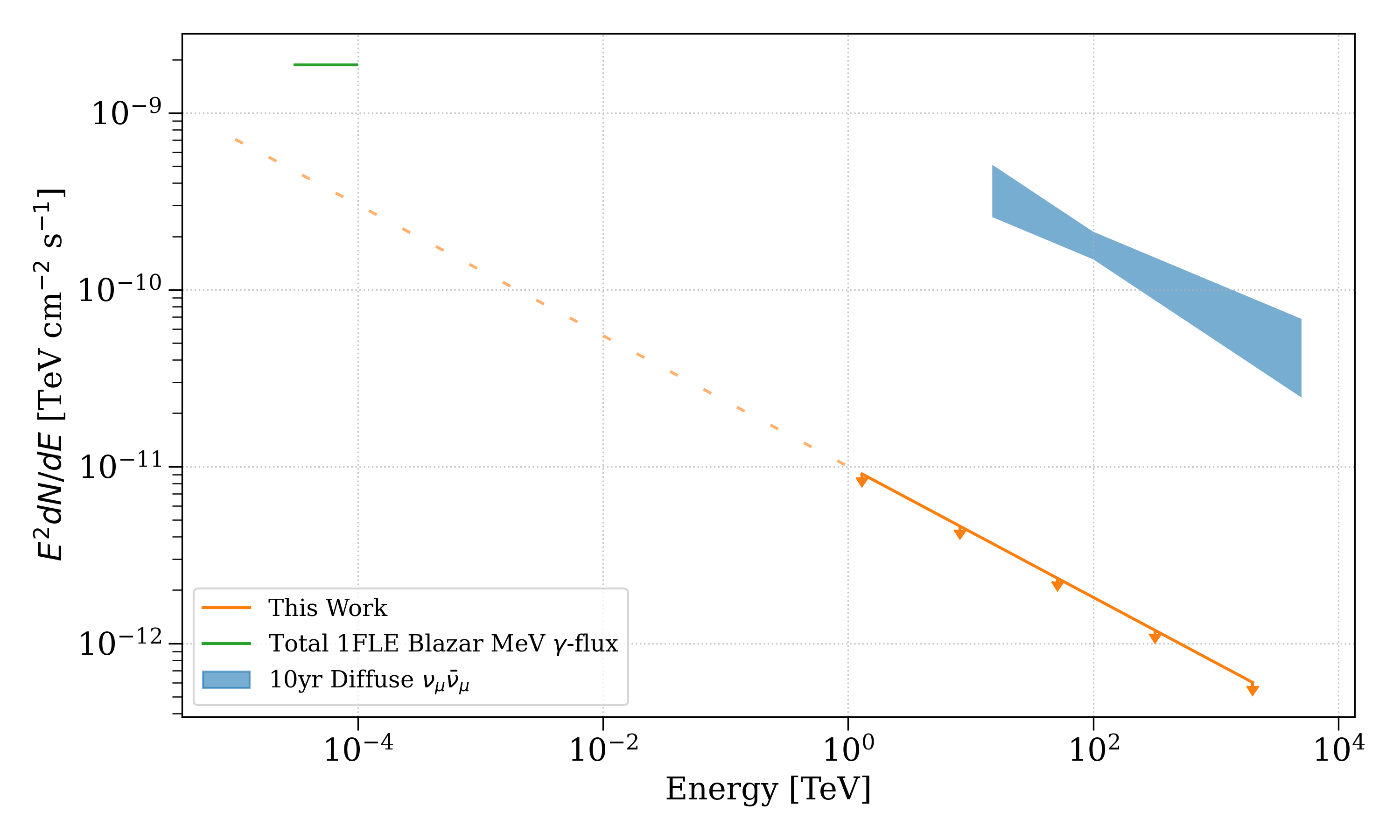}
    \caption{90\% confidence level upper limit on energy-scaled neutrino ($\nu_{\mu}+\bar{\nu}_{\mu}$) flux from 1FLE blazars assuming a simple power-law (orange) which is 1.0\% of the diffuse flux measurement from \citet{2022ApJ...928...50A} (blue, shown as a 68\% confidence level band). This upper limit does not include systematic uncertainties. The spectral index shown here, $\gamma=2.37$, is the best-fit of this diffuse measurement. The solid orange line shows the limit within the energy range which contributes 90\% of the total sensitivity. The dashed line extrapolates this limit to lower energies. The green line shows the sum of integrated gamma-ray fluxes between 30 and 100 MeV for 1FLE blazars which are used as source weights in obtaining the shown upper limit. Considering the relationship between this total flux and the limit from this analysis in the 30-100 MeV range, in conjunction with a gamma-ray model, could offer insight into the contribution of hadronic interactions to the observed blazar flux distribution. \label{fig:comp}}
\end{figure}

Figure \ref{fig:comp} compares the UL of this analysis under the flux-weighted source hypothesis to the astrophysical\deleted{, all-sky} diffuse $\nu_{\mu}+\bar{\nu}_{\mu}$ flux from \citet{2022ApJ...928...50A}. The UL of the analysis presented in this paper is on the per-flavor neutrino flux which for a track dataset, such as the one used here, can be closely approximated as a $\nu_{\mu}+\bar{\nu}_{\mu}$ flux. The spectral index of the UL shown in Figure \ref{fig:comp} is chosen to be 2.37 to match the best-fit spectral index of the compared diffuse measurement. By extending this limit to lower energies, it is also naively compared to the sum of 30-100 MeV photon energy fluxes from the 1FLE blazars used in this analysis. At around this spectral index and smaller, the hadronic component of the MeV gamma-ray flux may be constrained by the neutrino flux ULs presented here. The specifics of this constraint and its viability are dependent on the choice of blazar flux model but are beyond the scope of this paper.

\section{\label{sec:conc}Conclusions}
A stacking analysis searching for astrophysical neutrino production from 137 blazars in the \textit{Fermi}-LAT 1FLE catalog using 10 years of IceCube muon-neutrino data was presented. Observations were found, via the described methods, to be consistent with the background-only hypothesis for both equal and flux source weighting schemes. Upper limits were set on the energy-scaled neutrino flux from these blazars for both source weightings and various assumed spectral indices. Under the \added{30-100 MeV photon} flux \added{(equal)} weighting scheme and for a spectral index of $2.37$, the UL from this analysis is approximately 0.86 - 1.2\% \added{(1.3 - 1.8\%)} of the \added{most recent} diffuse $\nu_{\mu}+\bar{\nu}_{\mu}$ flux measurement\added{; this diffuse measurement uses IceCube track events observed in the northern celestial hemisphere from 2009 through 2018}. 

\added{The two blazar stacking analyses previously performed by IceCube and included in Table \ref{tab:comp} produced less stringent constraints on the diffuse astrophysical neutrino flux. It is important to consider, however, the differences between the analyses. In addition to the number of blazars and the energy range of the gamma-ray fluxes used to correlate with neutrino emission, each analysis uses a different set of IceCube data and compares to a different diffuse astrophysical neutrino flux. Each diffuse flux was measured using different sets of data and with varying methods; therefore, comparisons of constraints on these fluxes should be made with caution. \citet{2017ApJ...835...45A} uses IceCube data collected from the whole sky from 2009 to 2012. It reports an UL for 862 2LAC blazars with GeV gamma-ray flux weighting of 7\% of the diffuse flux from \citet{2015ApJ...809...98A}. The \citet{2015ApJ...809...98A} analysis performs a combined, all-sky fit of six diffuse flux measurements which include both cascade and track events from data-taking periods between 2009 and 2013. The best-fit spectral index from that analysis---which was also used to calculate the 2LAC UL---is 2.5. The analysis of 3FHL blazars \citep{2019ICRC...36..916H} compares its UL to the diffuse $\nu_{\mu}+\bar{\nu}_{\mu}$ flux from \citet{Haack:2017E1} which uses IceCube track events observed in the northern sky from 2009 through 2017; it finds a best-fit spectral index of 2.19. \citet{2019ICRC...36..916H} uses a sample of IceCube events from the northern hemisphere and reports an UL for 745 northern-sky, GeV-detected 3FHL blazars of 9.7 - 13.9\% of the diffuse flux for all blazar classes; the same analysis finds the upper limit for 101 FSRQs alone to be 2.9 - 3.8\% of the diffuse flux. }

\added{For a more direct comparison with \citet{2019ICRC...36..916H}, the UL from the work presented in this paper is compared to the \citet{Haack:2017E1} diffuse flux measurement. This UL for 137 1FLE blazars (of which 106 are FSRQs) with MeV gamma-ray flux weighting and an assumed spectral index of 2.19 is 1.1 - 1.9\% of the \citet{Haack:2017E1} diffuse flux. It is also worth noting that TXS 0506+056, which showed significant evidence for astrophysical neutrino emission \citep{2018Sci...361.1378I,2018Sci...361..147I}, is classified as a BL Lac, and its 30-100 MeV photon flux is roughly average in the 1FLE catalog. While it is possible that the 30-100 MeV photon energy range is not strongly correlated with emission of TeV-PeV astrophysical neutrinos in blazars, this analysis and the comparisons presented here cannot distinguish whether MeV or GeV photons have a stronger correlation, and further exploration is encouraged.}

It should be noted that this analysis contains limitations which result from the 1FLE catalog lying outside of \textit{Fermi}-LAT's most sensitive energy range. \added{This results in fewer detected objects.} \replaced{The}{Also, the} localization and flux precisions are worse than those in the most sensitive energies of this instrument, but are not accounted for with the methods of this analysis. Nevertheless, this analysis is a first search for astrophysical neutrino emission from 30-100 MeV gamma-ray-detected blazars and should lend itself as motivation for future studies into the low-energy gamma-ray band. Future electromagnetic observations in the MeV band will provide a larger number of sources with higher precision allowing for an updated analysis similar to the one presented here to be performed. For example, AMEGO-X \citep{Fleischhack:2021hJ} and e-ASTROGAM \citep{DEANGELIS20181} plan to offer much improved sensitivities in the MeV energy range and even down to a few hundred keV. Future analyses will also benefit from additional neutrino observations from upcoming improvements to neutrino observational capabilities, such as with IceCube-Gen2 \citep{Aartsen_2021G2}.

\section*{Acknowledgements}
The IceCube collaboration acknowledges the significant contributions to this manuscript from Michael A. Campana. 
USA {\textendash} U.S. National Science Foundation-Office of Polar Programs,
U.S. National Science Foundation-Physics Division,
U.S. National Science Foundation-EPSCoR,
Wisconsin Alumni Research Foundation,
Center for High Throughput Computing (CHTC) at the University of Wisconsin{\textendash}Madison,
Open Science Grid (OSG),
Extreme Science and Engineering Discovery Environment (XSEDE),
Frontera computing project at the Texas Advanced Computing Center,
U.S. Department of Energy-National Energy Research Scientific Computing Center,
Particle astrophysics research computing center at the University of Maryland,
Institute for Cyber-Enabled Research at Michigan State University,
and Astroparticle physics computational facility at Marquette University;
Belgium {\textendash} Funds for Scientific Research (FRS-FNRS and FWO),
FWO Odysseus and Big Science programmes,
and Belgian Federal Science Policy Office (Belspo);
Germany {\textendash} Bundesministerium f{\"u}r Bildung und Forschung (BMBF),
Deutsche Forschungsgemeinschaft (DFG),
Helmholtz Alliance for Astroparticle Physics (HAP),
Initiative and Networking Fund of the Helmholtz Association,
Deutsches Elektronen Synchrotron (DESY),
and High Performance Computing cluster of the RWTH Aachen;
Sweden {\textendash} Swedish Research Council,
Swedish Polar Research Secretariat,
Swedish National Infrastructure for Computing (SNIC),
and Knut and Alice Wallenberg Foundation;
Australia {\textendash} Australian Research Council;
Canada {\textendash} Natural Sciences and Engineering Research Council of Canada,
Calcul Qu{\'e}bec, Compute Ontario, Canada Foundation for Innovation, WestGrid, and Compute Canada;
Denmark {\textendash} Villum Fonden and Carlsberg Foundation;
New Zealand {\textendash} Marsden Fund;
Japan {\textendash} Japan Society for Promotion of Science (JSPS)
and Institute for Global Prominent Research (IGPR) of Chiba University;
Korea {\textendash} National Research Foundation of Korea (NRF);
Switzerland {\textendash} Swiss National Science Foundation (SNSF);
United Kingdom {\textendash} Department of Physics, University of Oxford.

\appendix
\section{Source List\label{app:sl}}
\startlongtable
\begin{deluxetable*}{llccccc}
    \tablenum{3}
    \tablecolumns{7}
    \tablecaption{List of blazars from the 1FLE catalog which are used in this analysis. \label{tab:sl}}
    \tablewidth{\textwidth}
    \tablehead{
    \colhead{Name} & \colhead{4FGL Association} & \colhead{Class} & \colhead{Right Ascension} & \colhead{Declination} & \colhead{30-100 MeV Energy Flux} & \colhead{Redshift$^a$} \\
    \colhead{} & \colhead{} & \colhead{} & \colhead{($^{\circ}$)} & \colhead{($^{\circ}$)} & \colhead{($10^{-12}$ erg cm$^{-2}$ s$^{-1}$)} & \colhead{}
    }
    \startdata
        1FLE J0003+2114 & 4FGL J0001.5+2113 & fsrq & 0.8 & 21.2 & 8.2 $\pm$ 2.7 & 1.11 \\  
        1FLE J0036-4236 & 4FGL J0030.3-4224 & fsrq & 9.2 & -42.6 & 8.3 $\pm$ 2.8 & 0.5 \\  
        1FLE J0109+0136 & 4FGL J0108.6+0134 & fsrq & 17.3 & 1.6 & 34 $\pm$ 10 & 2.1 \\  
        1FLE J0105+6233 & 4FGL J0109.7+6133 & fsrq & 16.5 & 62.6 & 35 $\pm$ 10 & 0.78 \\  
        1FLE J0111+3207 & 4FGL J0112.8+3208 & fsrq & 18.0 & 32.1 & 10.5 $\pm$ 3.5 & 0.6 \\  
        1FLE J0117+2420 & 4FGL J0115.8+2519 & bll & 19.4 & 24.3 & 10.4 $\pm$ 3.4 & 0.36 \\  
        1FLE J0119-2300 & 4FGL J0118.9-2141 & fsrq & 19.8 & -23.0 & 8.5 $\pm$ 2.8 & 1.16 \\  
        1FLE J0136+4752 & 4FGL J0137.0+4751 & fsrq & 24.2 & 47.9 & 13.8 $\pm$ 4.6 & 0.86 \\  
        1FLE J0144-2728 & 4FGL J0145.0-2732 & fsrq & 26.2 & -27.5 & 8.4 $\pm$ 2.8 & 1.15 \\  
        1FLE J0206-1705 & 4FGL J0205.0-1700 & fsrq & 31.6 & -17.1 & 7.8 $\pm$ 2.6 & 1.74 \\  
        1FLE J0213-5047 & 4FGL J0210.7-5101 & fsrq & 33.4 & -50.8 & 19.1 $\pm$ 5.8 & 1.0 \\  
        1FLE J0215+1034 & 4FGL J0211.2+1051 & bll & 33.8 & 10.6 & 7.4 $\pm$ 2.4 & -- \\  
        1FLE J0221+7403 & 4FGL J0217.4+7352 & fsrq & 35.4 & 74.1 & 18.9 $\pm$ 5.7 & 2.37 \\  
        1FLE J0213+0107 & 4FGL J0217.8+0144 & fsrq & 33.3 & 1.1 & 5.8 $\pm$ 1.9 & 1.72 \\  
        1FLE J0221+3604 & 4FGL J0221.1+3556 & fsrq & 35.4 & 36.1 & 26.3 $\pm$ 7.9 & 0.94 \\  
        1FLE J0219+4256 & 4FGL J0222.6+4302 & bll & 34.8 & 42.9 & 28.2 $\pm$ 8.5 & 0.44 \\  
        1FLE J0238+2900 & 4FGL J0237.8+2848 & fsrq & 39.6 & 29.0 & 29.4 $\pm$ 8.9 & 1.21 \\  
        1FLE J0238+1638 & 4FGL J0238.6+1637 & bll & 39.5 & 16.6 & 17.3 $\pm$ 5.2 & 0.94 \\  
        1FLE J0246-4621 & 4FGL J0245.9-4650 & fsrq & 41.7 & -46.4 & 13.6 $\pm$ 4.5 & 1.38 \\  
        1FLE J0254-2221 & 4FGL J0252.8-2219 & fsrq & 43.6 & -22.4 & 13.8 $\pm$ 4.6 & 1.43 \\  
        1FLE J0309+1002 & 4FGL J0309.0+1029 & fsrq & 47.5 & 10.0 & 6.6 $\pm$ 2.2 & 0.86 \\  
        1FLE J0304-6121 & 4FGL J0309.9-6058 & fsrq & 46.1 & -61.4 & 14.3 $\pm$ 4.7 & 1.48 \\  
        1FLE J0325+2204 & 4FGL J0325.7+2225 & fsrq & 51.5 & 22.1 & 11.0 $\pm$ 3.7 & 2.07 \\  
        1FLE J0329-3724 & 4FGL J0334.2-3725 & bll & 52.3 & -37.4 & 13.4 $\pm$ 4.4 & -- \\  
        1FLE J0331-3909 & 4FGL J0334.2-4008 & bll & 52.9 & -39.2 & 7.8 $\pm$ 2.6 & 1.36 \\  
        1FLE J0339-0133 & 4FGL J0339.5-0146 & fsrq & 54.8 & -1.6 & 13.9 $\pm$ 4.6 & 0.85 \\  
        1FLE J0349-2045 & 4FGL J0349.8-2103 & fsrq & 57.4 & -20.8 & 7.2 $\pm$ 2.4 & 2.94 \\  
        1FLE J0403-3554 & 4FGL J0403.9-3605 & fsrq & 60.8 & -35.9 & 40.8 $\pm$ 9.1 & 1.42 \\  
        1FLE J0424-0042 & 4FGL J0423.3-0120 & fsrq & 66.1 & -0.7 & 7.6 $\pm$ 2.5 & 0.92 \\  
        1FLE J0443-0024 & 4FGL J0442.6-0017 & fsrq & 71.0 & -0.4 & 7.6 $\pm$ 2.5 & 0.84 \\  
        1FLE J0448-4535 & 4FGL J0455.7-4617 & fsrq & 72.2 & -45.6 & 14.6 $\pm$ 4.8 & 0.86 \\  
        1FLE J0456-2322 & 4FGL J0457.0-2324 & fsrq & 74.2 & -23.4 & 30.3 $\pm$ 9.1 & 1.0 \\  
        1FLE J0500-0209 & 4FGL J0501.2-0158 & fsrq & 75.2 & -2.2 & 10.1 $\pm$ 3.3 & 2.29 \\  
        1FLE J0506+0528 $^b$ & 4FGL J0509.4+0542 & bll & 76.7 & 5.5 & 16.9 $\pm$ 5.6 & -- \\  
        1FLE J0531+0707 & 4FGL J0532.6+0732 & fsrq & 82.8 & 7.1 & 8.0 $\pm$ 2.6 & 1.25 \\  
        1FLE J0538-4438 & 4FGL J0538.8-4405 & bll & 84.5 & -44.6 & 32.7 $\pm$ 9.9 & 0.89 \\  
        1FLE J0649+4529 & 4FGL J0654.4+4514 & fsrq & 102.4 & 45.5 & 12.8 $\pm$ 4.2 & 0.93 \\  
        1FLE J0721+7121 & 4FGL J0721.9+7120 & bll & 110.4 & 71.4 & 46 $\pm$ 10 & 0.13 \\  
        1FLE J0729-1230 & 4FGL J0730.3-1141 & fsrq & 112.4 & -12.5 & 22.9 $\pm$ 6.9 & 1.59 \\  
        1FLE J0738+0133 & 4FGL J0739.2+0137 & fsrq & 114.5 & 1.6 & 15.9 $\pm$ 5.3 & 0.19 \\  
        1FLE J0805-0743 & 4FGL J0808.2-0751 & fsrq & 121.4 & -7.7 & 12.5 $\pm$ 4.1 & 1.84 \\  
        1FLE J0818+4202 & 4FGL J0818.2+4222 & bll & 124.7 & 42.0 & 14.5 $\pm$ 4.8 & 0.53 \\  
        1FLE J0823-2228 & 4FGL J0825.9-2230 & bll & 125.9 & -22.5 & 13.1 $\pm$ 4.3 & 0.91 \\  
        1FLE J0827+2440 & 4FGL J0830.8+2410 & fsrq & 126.8 & 24.7 & 11.4 $\pm$ 3.8 & 0.94 \\  
        1FLE J0841+7056 & 4FGL J0841.3+7053 & fsrq & 130.4 & 70.9 & 54 $\pm$ 12 & 2.22 \\  
        1FLE J0852-1149 & 4FGL J0850.1-1212 & fsrq & 133.2 & -11.8 & 7.3 $\pm$ 2.4 & 0.57 \\  
        1FLE J0855+2014 & 4FGL J0854.8+2006 & bll & 134.0 & 20.2 & 13.4 $\pm$ 4.4 & 0.31 \\  
        1FLE J0910+0159 & 4FGL J0909.1+0121 & fsrq & 137.6 & 2.0 & 5.9 $\pm$ 2.0 & 1.02 \\  
        1FLE J0921+4425 & 4FGL J0920.9+4441 & fsrq & 140.4 & 44.4 & 15.2 $\pm$ 5.0 & 2.19 \\  
        1FLE J0957+5503 & 4FGL J0957.6+5523 & fsrq & 149.3 & 55.1 & 8.8 $\pm$ 2.9 & 0.9 \\  
        1FLE J1005-2205 & 4FGL J1006.7-2159 & fsrq & 151.3 & -22.1 & 7.6 $\pm$ 2.5 & 0.33 \\  
        1FLE J1008-3201 & 4FGL J1008.8-3139 & bll & 152.0 & -32.0 & 7.7 $\pm$ 2.5 & -- \\  
        1FLE J1011+2442 & 4FGL J1012.7+2439 & fsrq & 152.8 & 24.7 & 6.6 $\pm$ 2.2 & 1.8 \\  
        1FLE J1031+6001 & 4FGL J1031.6+6019 & fsrq & 158.0 & 60.0 & 17.7 $\pm$ 5.3 & 1.23 \\  
        1FLE J1030+3811 & 4FGL J1032.6+3737 & bll & 157.6 & 38.2 & 7.0 $\pm$ 2.3 & -- \\  
        1FLE J1047+7131 & 4FGL J1048.4+7143 & fsrq & 161.9 & 71.5 & 51 $\pm$ 11 & 1.15 \\  
        1FLE J1049+2227 & 4FGL J1054.5+2211 & bll & 162.3 & 22.5 & 6.2 $\pm$ 2.1 & -- \\  
        1FLE J1059+0208 & 4FGL J1058.4+0133 & bll & 164.9 & 2.1 & 13.2 $\pm$ 4.4 & 0.89 \\  
        1FLE J1100+8117 & 4FGL J1058.5+8115 & fsrq & 165.1 & 81.3 & 17.7 $\pm$ 5.3 & 0.71 \\  
        1FLE J1105+3807 & 4FGL J1104.4+3812 & bll & 166.3 & 38.1 & 24.4 $\pm$ 7.4 & 0.03 \\  
        1FLE J1118-0558 & 4FGL J1121.4-0553 & fsrq & 169.7 & -6.0 & 6.1 $\pm$ 2.0 & 1.3 \\  
        1FLE J1125-1907 & 4FGL J1127.0-1857 & fsrq & 171.4 & -19.1 & 17.7 $\pm$ 5.3 & 1.05 \\  
        1FLE J1125+3639 & 4FGL J1127.8+3618 & fsrq & 171.4 & 36.7 & 13.1 $\pm$ 4.3 & 0.88 \\  
        1FLE J1138+3832 & 4FGL J1131.0+3815 & fsrq & 174.7 & 38.5 & 13.4 $\pm$ 4.4 & 1.73 \\  
        1FLE J1145+4001 & 4FGL J1146.9+3958 & fsrq & 176.5 & 40.0 & 21.6 $\pm$ 6.5 & 1.09 \\  
        1FLE J1148-3800 & 4FGL J1147.0-3812 & bll & 177.0 & -38.0 & 7.2 $\pm$ 2.4 & 1.05 \\  
        1FLE J1147+4836 & 4FGL J1153.4+4931 & fsrq & 176.9 & 48.6 & 7.8 $\pm$ 2.6 & 0.33 \\  
        1FLE J1201+2934 & 4FGL J1159.5+2914 & fsrq & 180.3 & 29.6 & 15.4 $\pm$ 5.1 & 0.72 \\  
        1FLE J1206+5421 & 4FGL J1208.9+5441 & fsrq & 181.6 & 54.4 & 10.1 $\pm$ 3.4 & 1.34 \\  
        1FLE J1219+2943 & 4FGL J1217.9+3007 & bll & 184.9 & 29.7 & 11.9 $\pm$ 3.9 & 0.13 \\  
        1FLE J1216+5019 & 4FGL J1223.9+5000 & fsrq & 184.1 & 50.3 & 8.4 $\pm$ 2.8 & 1.07 \\  
        1FLE J1224+2118 & 4FGL J1224.9+2122 & fsrq & 186.2 & 21.3 & 56 $\pm$ 13 & 0.44 \\  
        1FLE J1227+0218 & 4FGL J1229.0+0202 & fsrq & 187.0 & 2.3 & 65 $\pm$ 14 & 0.16 \\  
        1FLE J1247-2541 & 4FGL J1246.7-2548 & fsrq & 191.8 & -25.7 & 20.5 $\pm$ 6.2 & 0.64 \\  
        1FLE J1256-0545 & 4FGL J1256.1-0547 & fsrq & 194.0 & -5.8 & 63 $\pm$ 14 & 0.54 \\  
        1FLE J1256-2314 & 4FGL J1258.8-2219 & fsrq & 194.1 & -23.2 & 13.4 $\pm$ 4.4 & 1.3 \\  
        1FLE J1309+3302 & 4FGL J1310.5+3221 & fsrq & 197.4 & 33.0 & 12.5 $\pm$ 4.1 & 1.0 \\  
        1FLE J1324+2247 & 4FGL J1321.1+2216 & fsrq & 201.1 & 22.8 & 6.0 $\pm$ 2.0 & 0.94 \\  
        1FLE J1332-0518 & 4FGL J1332.0-0509 & fsrq & 203.2 & -5.3 & 15.2 $\pm$ 5.0 & 2.15 \\  
        1FLE J1333-1250 & 4FGL J1332.6-1256 & fsrq & 203.4 & -12.8 & 16.0 $\pm$ 5.3 & 1.5 \\  
        1FLE J1342+6530 & 4FGL J1338.0+6534 & fsrq & 205.6 & 65.5 & 8.7 $\pm$ 2.9 & 0.95 \\  
        1FLE J1344+4456 & 4FGL J1345.5+4453 & fsrq & 206.2 & 44.9 & 27.3 $\pm$ 8.2 & 2.53 \\  
        1FLE J1416-0818 & 4FGL J1419.4-0838 & fsrq & 214.1 & -8.3 & 6.4 $\pm$ 2.1 & -- \\  
        1FLE J1427-4205 & 4FGL J1427.9-4206 & fsrq & 217.0 & -42.1 & 77 $\pm$ 17 & 1.55 \\  
        1FLE J1436+2409 & 4FGL J1436.9+2321 & fsrq & 219.2 & 24.2 & 14.3 $\pm$ 4.8 & 1.54 \\  
        1FLE J1456-3612 & 4FGL J1457.4-3539 & fsrq & 224.2 & -36.2 & 12.7 $\pm$ 4.2 & 1.42 \\  
        1FLE J1503+1033 & 4FGL J1504.4+1029 & fsrq & 225.9 & 10.6 & 31.7 $\pm$ 9.6 & 1.84 \\  
        1FLE J1512-0913 & 4FGL J1512.8-0906 & fsrq & 228.2 & -9.2 & 156 $\pm$ 32 & 0.36 \\  
        1FLE J1518-2422 & 4FGL J1517.7-2422 & bll & 229.7 & -24.4 & 14.7 $\pm$ 4.9 & 0.05 \\  
        1FLE J1522+3147 & 4FGL J1522.1+3144 & fsrq & 230.6 & 31.8 & 56 $\pm$ 12 & 1.49 \\  
        1FLE J1518-2650 & 4FGL J1522.6-2730 & bll & 229.5 & -26.8 & 10.3 $\pm$ 3.4 & 1.29 \\  
        1FLE J1559+1149 & 4FGL J1555.7+1111 & bll & 239.9 & 11.8 & 7.8 $\pm$ 2.7 & -- \\  
        1FLE J1603+5734 & 4FGL J1604.6+5714 & fsrq & 240.8 & 57.6 & 12.2 $\pm$ 4.1 & 0.72 \\  
        1FLE J1626-7732 & 4FGL J1617.9-7718 & fsrq & 246.7 & -77.5 & 12.7 $\pm$ 4.2 & 1.71 \\  
        1FLE J1625-2510 & 4FGL J1625.7-2527 & fsrq & 246.4 & -25.2 & 38 $\pm$ 12 & 0.79 \\  
        1FLE J1625-2926 & 4FGL J1626.0-2950 & fsrq & 246.4 & -29.4 & 12.1 $\pm$ 4.0 & 0.82 \\  
        1FLE J1636+3831 & 4FGL J1635.2+3808 & fsrq & 249.1 & 38.5 & 73 $\pm$ 16 & 1.81 \\  
        1FLE J1703+6815 & 4FGL J1700.0+6830 & fsrq & 255.8 & 68.3 & 14.1 $\pm$ 4.8 & 0.3 \\  
        1FLE J1720+0916 & 4FGL J1722.7+1014 & fsrq & 260.1 & 9.3 & 8.0 $\pm$ 2.7 & 0.73 \\  
        1FLE J1739+5134 & 4FGL J1740.5+5211 & fsrq & 264.8 & 51.6 & 11.7 $\pm$ 3.9 & 1.38 \\  
        1FLE J1735+7015 & 4FGL J1748.6+7005 & bll & 263.9 & 70.3 & 16.6 $\pm$ 5.5 & 0.77 \\  
        1FLE J1753+7815 & 4FGL J1800.6+7828 & bll & 268.4 & 78.3 & 20.4 $\pm$ 6.2 & 0.68 \\  
        1FLE J1827+6859 & 4FGL J1823.5+6858 & bll & 276.9 & 69.0 & 23.3 $\pm$ 7.0 & -- \\  
        1FLE J1831-5805 & 4FGL J1832.6-5658 & bll & 277.9 & -58.1 & 21.2 $\pm$ 6.4 & -- \\  
        1FLE J1834-2116 & 4FGL J1833.6-2103 & fsrq & 278.6 & -21.3 & 23.4 $\pm$ 7.1 & 2.51 \\  
        1FLE J1838+6812 & 4FGL J1842.3+6810 & fsrq & 279.6 & 68.2 & 20.8 $\pm$ 6.3 & 0.47 \\  
        1FLE J1845+3238 & 4FGL J1848.4+3217 & fsrq & 281.3 & 32.6 & 16.5 $\pm$ 5.5 & 0.8 \\  
        1FLE J1914-2017 & 4FGL J1911.2-2006 & fsrq & 288.5 & -20.3 & 14.2 $\pm$ 4.7 & 1.12 \\  
        1FLE J1938-6215 & 4FGL J1941.3-6210 & bll & 294.5 & -62.3 & 17.5 $\pm$ 5.3 & -- \\  
        1FLE J1958-3919 & 4FGL J1958.0-3845 & fsrq & 299.6 & -39.3 & 16.1 $\pm$ 5.3 & 0.63 \\  
        1FLE J2001+6556 & 4FGL J2007.2+6607 & fsrq & 300.3 & 65.9 & 14.1 $\pm$ 4.7 & 1.32 \\  
        1FLE J2028+7651 & 4FGL J2022.5+7612 & bll & 307.2 & 76.9 & 16.4 $\pm$ 5.4 & 0.59 \\  
        1FLE J2027-0805 & 4FGL J2025.6-0735 & fsrq & 306.8 & -8.1 & 21.1 $\pm$ 6.4 & 1.39 \\  
        1FLE J2035+1102 & 4FGL J2035.4+1056 & fsrq & 308.8 & 11.0 & 17.6 $\pm$ 5.3 & 0.6 \\  
        1FLE J2039+5042 & 4FGL J2038.7+5117 & fsrq & 309.9 & 50.7 & 21.5 $\pm$ 6.5 & 1.69 \\  
        1FLE J2056-4724 & 4FGL J2056.2-4714 & fsrq & 314.2 & -47.4 & 31.3 $\pm$ 9.4 & 1.49 \\  
        1FLE J2120-4549 & 4FGL J2126.3-4605 & fsrq & 320.2 & -45.8 & 9.8 $\pm$ 3.3 & 1.67 \\  
        1FLE J2144+1751 & 4FGL J2143.5+1743 & fsrq & 326.2 & 17.9 & 22.1 $\pm$ 6.7 & 0.21 \\  
        1FLE J2141-7616 & 4FGL J2147.3-7536 & fsrq & 325.4 & -76.3 & 19.9 $\pm$ 6.0 & 1.14 \\  
        1FLE J2155-3018 & 4FGL J2151.8-3027 & fsrq & 328.9 & -30.3 & 39 $\pm$ 12 & 2.35 \\  
        1FLE J2209-8335 & 4FGL J2201.5-8339 & fsrq & 332.4 & -83.6 & 11.5 $\pm$ 3.8 & 1.86 \\  
        1FLE J2156+5102 & 4FGL J2201.8+5048 & fsrq & 329.2 & 51.0 & 27.5 $\pm$ 8.3 & 1.9 \\  
        1FLE J2203+4214 & 4FGL J2202.7+4216 & bll & 330.8 & 42.2 & 57 $\pm$ 13 & 0.07 \\  
        1FLE J2227-0838 & 4FGL J2229.7-0832 & fsrq & 337.0 & -8.6 & 18.7 $\pm$ 5.6 & 1.56 \\  
        1FLE J2231+1132 & 4FGL J2232.6+1143 & fsrq & 337.9 & 11.5 & 69 $\pm$ 15 & 1.04 \\  
        1FLE J2232-4923 & 4FGL J2235.3-4836 & fsrq & 338.0 & -49.4 & 6.8 $\pm$ 2.3 & 0.51 \\  
        1FLE J2236+2807 & 4FGL J2236.3+2828 & fsrq & 339.2 & 28.1 & 9.2 $\pm$ 3.0 & 0.79 \\  
        1FLE J2240-1340 & 4FGL J2236.5-1433 & bll & 340.1 & -13.7 & 8.3 $\pm$ 2.7 & -- \\  
        1FLE J2251+4034 & 4FGL J2244.2+4057 & fsrq & 342.8 & 40.6 & 9.8 $\pm$ 3.3 & 1.17 \\  
        1FLE J2256-2750 & 4FGL J2250.7-2806 & bll & 344.1 & -27.8 & 22.4 $\pm$ 6.8 & 0.52 \\  
        1FLE J2254+1617 & 4FGL J2253.9+1609 & fsrq & 343.6 & 16.3 & 273 $\pm$ 56 & 0.86 \\  
        1FLE J2311+3404 & 4FGL J2311.0+3425 & fsrq & 348.0 & 34.1 & 13.3 $\pm$ 4.4 & 1.82 \\  
        1FLE J2320-0409 & 4FGL J2323.5-0317 & fsrq & 350.0 & -4.2 & 8.6 $\pm$ 2.9 & 1.39 \\  
        1FLE J2329+0858 & 4FGL J2327.5+0939 & fsrq & 352.5 & 9.0 & 10.6 $\pm$ 3.5 & 1.84 \\  
        1FLE J2330-4033 & 4FGL J2328.3-4036 & fsrq & 352.5 & -40.6 & 7.7 $\pm$ 2.6 & -- \\  
        1FLE J2329-4929 & 4FGL J2329.3-4955 & fsrq & 352.4 & -49.5 & 37 $\pm$ 11 & 0.52 \\  
        1FLE J2345-1611 & 4FGL J2345.2-1555 & fsrq & 356.5 & -16.2 & 15.7 $\pm$ 5.2 & 0.62 \\   
    \enddata
    \tablecomments{The provided uncertainties on the energy flux measurements are $1\sigma$ errors. Classifications are obtained from the blazars' associations in the 4FGL-DR2 catalog. All other quantities obtained from the 1FLE catalog.\\
    $^a$Redshift in the 1FLE catalog is obtained from the sources' associations in the \textit{Fermi}-LAT 3LAC catalog. A redshift of ``--'' indicates no association or no redshift information. \\
    $^b$This is the same blazar discussed in Section \ref{sec:Intro}, TXS 0506+056.}
\end{deluxetable*}

\newpage
\bibliography{bib.bib}{}
\bibliographystyle{aasjournal}

\listofchanges
\end{document}